\documentclass[fleqn,usenatbib]{mnras}
\usepackage[T1]{fontenc}
\usepackage{ae,aecompl}
\usepackage{graphicx}
\usepackage{multicol}
\usepackage{float}

\usepackage{graphicx}	
\usepackage{amsmath}	
\usepackage{amssymb}	
\usepackage{threeparttable}
\usepackage{subfigure}
\usepackage{multirow}
\usepackage{multicol}
\usepackage{booktabs}
\usepackage{makecell}
\newcommand{\orcid}[1]{\href{https://orcid.org/#1}{\includegraphics[width=8pt]{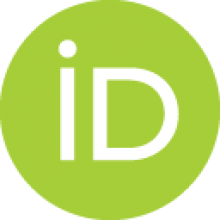}}}

\title[]{The Spin of New Black Hole Candidate: MAXI J1803-298 Observed by \emph{NuSTAR} and \emph{NICER}}

\author[Ye Feng et al.]{
Ye Feng\orcid{0000-0002-6961-8082}$^{1,2}$,
Xueshan Zhao\orcid{0000-0002-3281-5306}$^{1,2}$,
Yufeng Li\orcid{0000-0002-4135-074X}$^{1,2}$,
Lijun Gou\orcid{0000-0003-3057-5860}$^{1,2}$\thanks{E-mail: lgou@nao.cas.cn},
Nan Jia\orcid{0000-0002-5487-880X}$^{1,2}$,
\newauthor Zhenxuan Liao\orcid{0000-0003-2221-0490}$^{1,2}$,
Yuan Wang\orcid{0000-0001-9604-0370}$^{1,2}$
\\
$^{1}$Key Laboratory for Computational Astrophysics, National Astronomical Observatories, Chinese Academy of Sciences, \\Datun Road A20, Beijing 100012, China\\
$^{2}$School of Astronomy and Space Sciences, University of Chinese Academy of Sciences, Datun Road A20, Beijing 100049, China\\
}

\date{Accepted XXX. Received YYY; in original form ZZZ}

\pubyear{2021}

\begin{document}
\pagerange{\pageref{firstpage}--\pageref{lastpage}}
\maketitle

\begin{abstract}
MAXI J1803-298, a newly-discovered Galactic transient, and black hole candidate was first detected by \emph{MAXI}/GSC on May 1st, 2021. In this paper, we present a detailed spectral analysis of MAXI J1803-298. Utilizing the X-ray reflection fitting method, we perform a joint fit to the spectra of MAXI J1803-298, respectively, observed by \emph{NuSTAR} and \emph{NICER}/XTI on the same day over the energy range between 0.7-79.0 keV in SIMS state, and the observed inner radius of the accretion disk is demonstrated to have extended to the ISCO radius ($R_{\text{in}}= $1.08 $R_{\text{ISCO}}$). Implementing the relativistic reflection model \texttt{relxillCp}, we found its spin (and the inclination angle $i$) can be constrained to be close to an extreme value, 0.991 ($i\sim$ $70 ^{\circ}$), at 68\% confidence interval. The results suggest that MAXI J1803-298 may be a fast-rotating black hole with a large inclination angle.  
\end{abstract}


\section{Introduction}
\label{section:1}
Black holes (BHs) are the simplest objects in the universe, according to the theory of general relativity (GR). Mass, and angular momentum (also known as spin) can completely characterize an astrophysical black hole. Among the two parameters, spin is crucial for understanding black hole physics and jet power (\citealt{Narayan2012}). It is usually defined as $a_{*} = cJ/GM^2$ mathematically, where $J$ is the angular momentum of the black hole, $c$ is the speed of light, $M$ is the black hole mass and $G$ is the gravitational constant. The dimensionless spin lies in a range from -1 to 1. For a Schwarzschild black hole, its spin is 0, while for a Kerr black hole, its spin is $\pm$ 1 ($\pm$ here only means that a black hole is prograde or retrograde).

The X-ray reflection fitting method (also called as $\text {Fe K} \alpha$ fitting method) is widely used in estimating the spin of black hole in X-ray binary (e.g., \citealt{Miller2013, Draghis2020, Feng2021}), which by means of modeling the relativistic reflection component in a spectrum (see \citealt{Fabian1989} for details). 

Generally, an outbursting black hole X-ray binary (BHXRB) usually presents four spectral states in turn, low hard state (LHS), hard intermediate state (HIMS), soft intermediate state (SIMS, also named as steep power-law (SPL)), and high soft state (HSS, also called as thermal dominant (TD)), as the q-like X-ray hardness-intensity diagram (HID) shows (see \citealt{McClintock2006} for details). Typically, the photon index of a spectrum is 1.4-1.6 for LHS, 1.6-2.5 for HIMS/SIMS and 2.5-4.0 for HSS (\citealt{Belloni2006}).

MAXI J1803-298 is a new X-ray binary system hosting a black-hole candidate which was detected in outburst by the Gas Slit Camera of the \emph{Monitor of All-sky X-ray Image} (\emph{MAXI}/GSC) nova alert system at 19:50 UT on May 1st, 2021, which locates at R.A.=$270.923^{\circ}$, Dec=$-29.804^{\circ}$ (J2000) (\citealt{Serino2021}). On May 4th (MJD 59338.9), MAXI J1803-298 was observed for 15 minutes on-target at a central frequency of 1.284 GHz (L band) with a total bandwidth of 860 MHz by \emph{Meer Karoo Array Telescope} (\emph{MeerKAT}) (\citealt{Espinasse2021}). Later, other telescopes such as \emph{the Neutron star Interior Composition Explorer} (\emph{NICER}) (\citealt{Gendreau2021, Bult2021}), \emph{Neil Gehrels Swift Observatory} (\emph{Swift}) (\citealt{Gropp2021}), \emph{Multicolor Imaging Telescopes for Survey and Monstrous Explosions} (\emph{MITSuME}) (\citealt{Hosokawa2021}), \emph{the Southern African Large Telescope} (\emph{SALT}) (\citealt{Buckley2021}), \emph{International Gamma-Ray Astrophysics Laboratory} (\emph{INTEGRAL}) (\citealt{Chenevez2021}), \emph{Nuclear Spectroscopic Telescope Array} (\emph{NuSTAR}) (\citealt{Xu2021}), \emph{the Hard X-ray Modulation Telescope} (\emph{HXMT}) (\citealt{Wang2021}) and \emph{Astronomy Satellite} (\emph{AstroSAT}) (\citealt{Jana2021}) conducted follow-up observations at multiple energy bands. The transition of X-ray spectral states was observed in the following months after discovery (\citealt{Steiner2021, Shidatsu2021}). \emph{Swift} (\citealt{Miller2021}), \emph{NuSTAR} (\citealt{Xu2021}) and \emph{NICER} (\citealt{Homan2021}) observed strong periodic X-ray absorption dips in MAXI J1803-298, which demonstrate that MAXI J1803-298 has a large viewing angle within $60^{\circ}-80^{\circ}$. In addition, \emph{NICER} (\citealt{Ubach2021}), \emph{AstroSat} (\citealt{Chand2021}) and \emph{HXMT} (\citealt{Wang2021}) have detected quasi-periodic oscillations (QPOs), respectively. 
 
Based on the X-ray reflection fitting method, we jointly fit the spectra of MAXI J1803-298 observed by \emph{NuSTAR} and \emph{NICER}/XTI on the same day over the energy range of 0.7-79.0 keV. Based on the reflection features of the new black hole candidate MAXI J1803-298, we constrain its spin and inclination angle.

The paper is organized as follows. We introduce the observations and data reduction about \emph{NuSTAR} and \emph{NICER}/XTI in Section \ref{section:2}. Then, we demonstrate the spectral analysis in detail in Section \ref{section:3}. In Section \ref{section:4} and Section \ref{section:5}, we present the discussions and conclusions, respectively.

\setcounter{figure}{0}
\begin{figure}
\subfigure{
    \label{fig:1_1}
    \includegraphics[angle=0, width=\columnwidth]{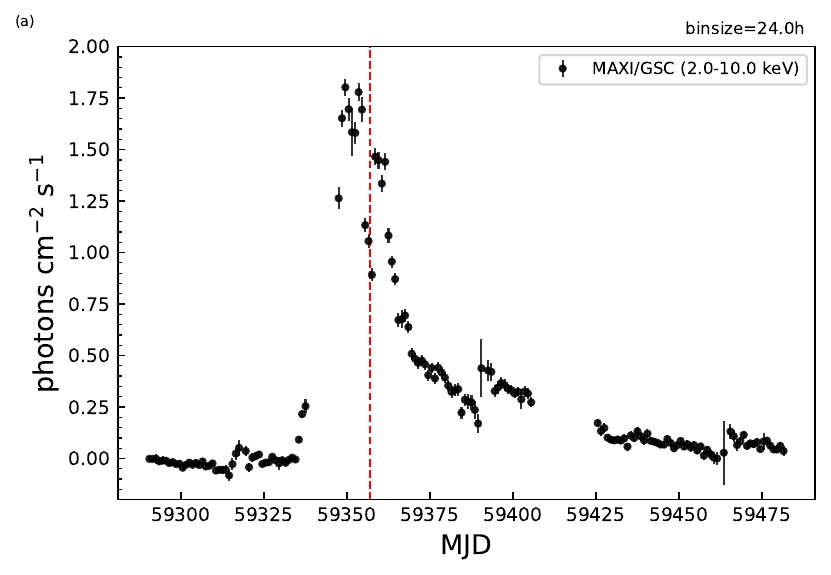}
    }
    \subfigure{
    \includegraphics[angle=0, width=\columnwidth]{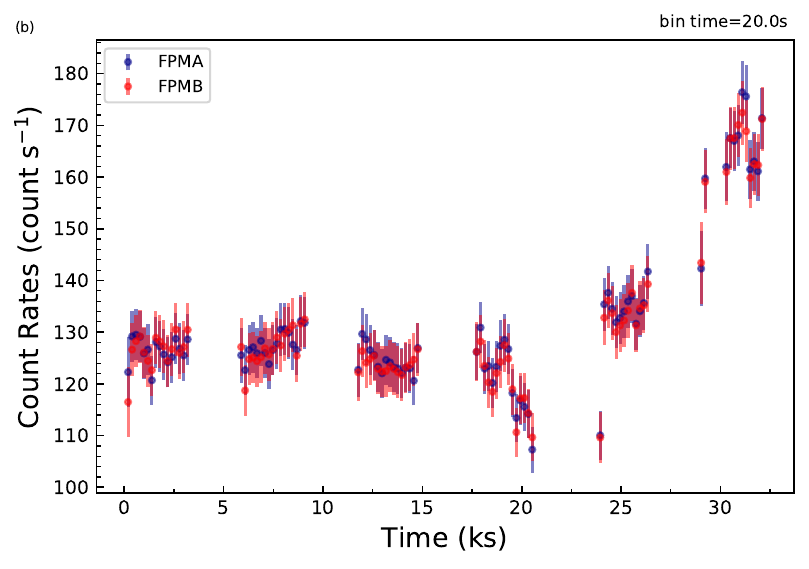}
      \label{fig:1_2}
    }
    \caption{(a) \emph{MAXI}/GSC observations of MAXI J1803-298 in 2.0-10.0 keV. The red dashed line represents the time observed by \emph{NuSTAR} (ObsID: 90702318002) and \emph{NICER} (ObsID: 4202130110). (b) Take the observation of \emph{NuSTAR} (ObsID: 90702318002) as an example, showing full band net light curve of MAXI J1803-298. Respectively, FPMA and FPMB data are plotted in blue and red.}
\end{figure}

\section{OBSERVATIONS AND DATA REDUCTION}
\label{section:2}
The outburst of the new X-ray transient MAXI J1803-298, \emph{NuSTAR} and the X-ray Timing Instrument payload on the \emph{NICER} (\emph{NICER}/XTI) were both observed on May 23rd, 2021. The observation data is marked with a red dashed line in the light curves of this source shown in Figure \ref{fig:1_1}. During this outburst of MAXI J1803-298, its net light curve in Figure \ref{fig:1_2} shows obvious variabilities, which may indicate that the system has a high inclination (\citealt{Frank1987}). In order to explore the reflection features of MAXI J1803-298, the data from \emph{NuSTAR} (\citealt{Harrison2013}) and \emph{NICER}/XTI (\citealt{Gendreau2012}) are both utilized. The advantage of joint fit to the data of these two detectors lies in that the hydrogen column density ($N_{\mathrm{H}}$) can be well constrained with the low energy observation from \emph{NICER}/XTI, and the reflection composition can be better fitted with the broad energy band of \emph{NuSTAR}. It is noted that neither observation was affected by the pile-up effect. 

In general, there should be no dips, flares, or other obvious fluctuations in the spectra used to measure the spin during this period. As a result, we utilize the commands \texttt{ftselect}\footnote{\url{https://heasarc.gsfc.nasa.gov/lheasoft/ftools/headas/ftselect.html}} and \texttt{ftdelrow}\footnote{\url{https://heasarc.gsfc.nasa.gov/lheasoft/ftools/headas/ftdelrow.html}} to remove the dip time from \emph{NuSTAR} and \emph{NICER}/XTI event data. The detailed observational information of spectra is listed in Table \ref{table:1}. In the process of spectral fitting, the \texttt{XSPEC v12.11.1} (\citealt{Arnaud1996}) software package is implemented with $\chi^2$ statistics. Using command \texttt{grppha}, all the data are grouped to achieve at least 30 photons per energy bin. Moreover, we also ignore all bad channels for all the spectra. Considering the cross-calibration between \emph{NuSTAR}/(FPMA|FPMB) and \emph{NICER}/XTI, we also make the Crab correction in \texttt{XSPEC} using the multiplicative model \texttt{crabcor} for \emph{NuSTAR} and \emph{NICER}/XTI (see \citealt{Steiner2010} for details).

\subsection{\emph{NuSTAR} OBSERVATION}
\label{section:2:1}
MAXI J1803-298 was observed by \emph{NuSTAR} on May 23, 2021 (ObsID: 90702318002). According to standard procedures\footnote{\url{https://heasarc.gsfc.nasa.gov/docs/nustar/analysis/nustar_swguide.pdf}}, we process the \emph{NuSTAR} data using HEAsoft v6.28 based on the calibration files v20210315. We use a circle centered on the position of the source with a radius of $80^{\prime \prime}$ to extract the source spectra. The background spectra are extracted with the same radius. In addition, we also make the dead-time correction. FPMA and FPMB are fitted jointly over 3.0-79.0 keV. 

\subsection{\emph{NICER} OBSERVATION}
\label{section:2:2}
Considering that \emph{NICER} has three observations on the same day with \emph{NuSTAR}, we select the observation (OBsID: 4202130110) with more overlapping time with \emph{NuSTAR} observation. We process the \emph{NICER} data following the standard steps\footnote{\url{https://heasarc.gsfc.nasa.gov/docs/nicer/analysis_threads/}} based on the latest calibration files\footnote{\url{https://heasarc.gsfc.nasa.gov/docs/heasarc/caldb/caldb_supported_missions.html}} v202107220315. We use tools \texttt{nicerl2}\footnote{\url{https://heasarc.gsfc.nasa.gov/docs/nicer/analysis_threads/nicerl2/}} and \texttt{nibackgen3C50}\footnote{\url{https://heasarc.gsfc.nasa.gov/docs/nicer/tools/nicer_bkg_est_tools.html}} to extract the source and background spectra, respectively. The response matrix 20170601v003 and ancillary file 20170601v005\footnote{\url{https://heasarc.gsfc.nasa.gov/docs/nicer/analysis_threads/arf-rmf/}} are used. Dead-time correction is applied. Considering the impact of low energy noise (< 0.5 keV), in this work, the spectrum of \emph{NICER} is fitted over the energy band of 0.7-10.0 keV, ignoring  the energy range of 1.7-2.1 keV (calibration residuals remain found in Si band) and 2.2-2.3 keV (calibration residuals remain in Au edges), respectively. Details of \emph{NICER} calibration are described in \cite{LaMarr2016}. Furthermore, following \cite{Jaisawal2019}, we also add the systematic error of 1.5\% with command \texttt{fparkey}\footnote{\url{https://heasarc.gsfc.nasa.gov/lheasoft/ftools/fhelp/fparkey.html}}.

\section{SPECTRAL ANALYSIS}
\label{section:3}

\subsection{PHENOMENOLOGICAL MODEL}
\label{section:3.1}
We first fit the \emph{NuSTAR} and \emph{NICER}/XTI data with our preliminary model: \texttt{crabcor*constant*TBabs*(diskbb+powerlaw)}. Model \texttt{crabcor} is described in Section \ref{section:2}. Model \texttt{constant} coordinates the calibration differences in different instruments in the process of joint fit. As to the model \texttt{constant}, we mainly fix the constant of \emph{NICER} $C_{\mathrm{NICER}}$ to 1 and allow the constant of FPMA $C_{\mathrm{FPMA}}$ and FPMB $C_{\mathrm{FPMB}}$ to vary. The subsequent usage of \texttt{constant} is the same. \texttt{TBabs} is the Galactic interstellar medium absorption model, and we set the abundance in \cite{Wilms2000} and put the cross-section in \cite{Verner1996}. Finally, we find a broadened iron line at 6.0-7.0 keV, and a Compton hump at 20.0-30.0 keV with a reduced chi-square $\chi_{v}^{2}=1.64$ (4244.66/2592), by ignoring the 6.0-8.0 keV and 20.0-40.0 keV energy range (Figure \ref{fig:2}). 

To study the reflection features of MAXI J1803-298, we firstly focus on the basic properties of the broadened iron line with model: \texttt{crabcor*constant*TBabs*(diskbb+powerlaw+gauss)}. After setting the central energy of the Gaussian line profile to 6.4-6.97 keV, we obtain that the equivalent width of the broadened iron line is 618 eV for \emph{NICER} data and 418 eV for \emph{NuSTAR} data, respectively, and the line peak is pegged at $E=6.97$ keV, with $\chi_{v}^{2}=1.42$ (3670.21/2589). 

\subsection{PHYSICAL MODEL}
\label{section:3.2}
In order to analyze the reflection composition better, we use a physical model which utilizes a Comptonization continuum as an incident spectrum, i.e.,  \texttt{crabcor*constant*TBabs*(diskbb+relxillCp)}. To ensure the inner radius of the accretion disk has extended to the innermost stable circular orbit (ISCO), during the fitting process, we set the disk inner radius ($R_{\text{in}}$) fitted freely, while we fix the dimensionless spin of black hole ($a_{*}$) to its maximum value ($a_{*}=0.998$). We also bound the emissivity index of inner region ($q_{\mathrm{in}}$) in 3-10, the emissivity index of outer region ($q_{\mathrm{out}}$) in the range of 0-3, and the break radius ($R_{\mathrm{br}}$) in $2R_{\mathrm{g}}-6R_{\mathrm{g}}$, where $R_{\mathrm{g}} = \mathrm{GM} / c^{2}$ is the gravitational radius (see \citealt{Miller2018} for details), then allow them to fit freely within the range above. Owing to the limited fitting energy range, we also fix the electron temperature of corona ($kT_{\text{e}}$) at 1/3 default value of the power-law cut-off energy ($E_{\text {cut}}=300 \mathrm{~keV}$), i.e., $kT_{\text{e}}= 100 \mathrm{~keV}$. The outer radius of disk ($R_{\text{out}}$) is fixed at its default value (400 $R_{\text{g}}$), iron abundance ($A_{\mathrm{Fe}}$) is fixed at a solar iron abundance, and redshift (z) is fixed at zero. We set other parameters (hydrogen column density ($N_{\mathrm{H}}$), inclination angle ($i$), photon index of the X-ray spectrum ($\Gamma$), ionization state of accretion disk ($\log \xi $), reflection fraction ($R_{\text{ref}}$) and normalization ($N_{\text {relxillCp}}$) ) to vary freely. We name the model above as Model A. We also define our baseline model, Model B, which differs only from Model A in freeing the spin and fixing the $R_{\text{in}}$ to an ISCO. We list the detailed fitting results for both Model A and B in Table \ref{table:2}, and the spectral fit of Model B is shown in Figure \ref{fig:3}. We also use command \texttt{steppar} in \texttt{XSPEC}, to search for the best fit for the spin from 0.99 to 0.998 for our baseline model, as shown in Figure \ref{fig:4}. The three dashed lines represent 99\%, 90\%, and 68\% confidence levels from top to bottom.

\setcounter{figure}{1}
\begin{figure}
    \includegraphics[angle=0, width=\columnwidth]{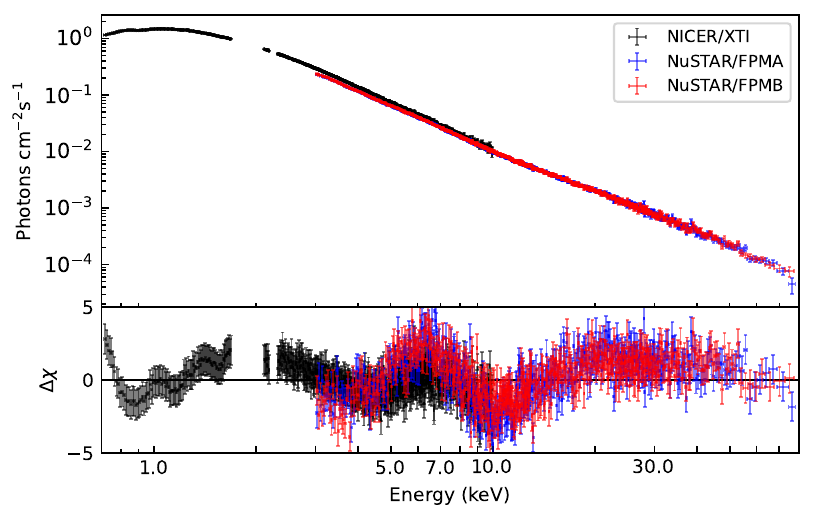}
    \caption{Using Model: \texttt{crabcor*constant*TBabs*(diskbb+powerlaw)}, it shows obvious relativistic reflection features. The data are fitted over 0.7-79.0 keV, ignoring 6.0-8.0 keV and 20.0-40.0 keV. \emph{NICER}/XTI, \emph{NuSTAR}/FPMA and \emph{NuSTAR}/FPMB data are plotted in black, blue and red, respectively. Data have been rebinned for visual clarity.}
        \label{fig:2}
\end{figure}

\setcounter{figure}{2}
\begin{figure}
    \includegraphics[angle=0, width=\columnwidth]{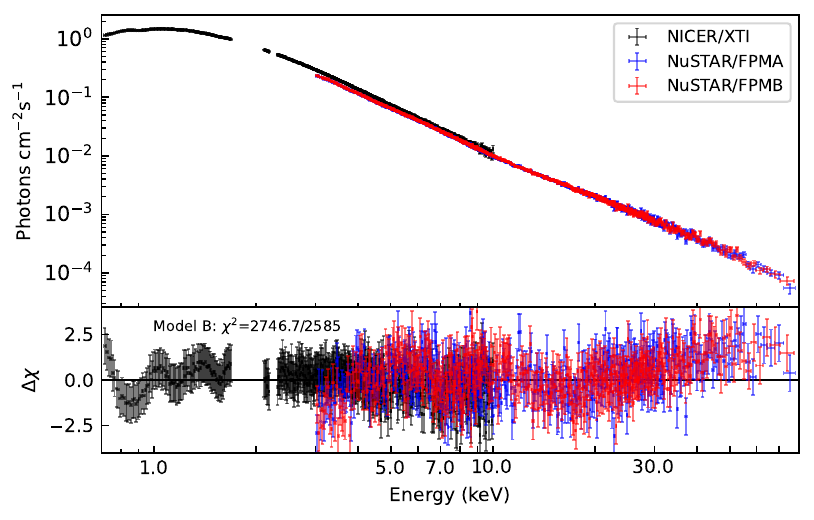}
    \caption{The fitting result of Model B. \emph{NICER}/XTI, \emph{NuSTAR}/FPMA and \emph{NuSTAR}/FPMB data are plotted in black, blue and red, respectively. Data have been rebinned for visual clarity.}
        \label{fig:3}
\end{figure}

\section{DISCUSSION}
\label{section:4}
\subsection{A COMPACT CORONA CLOSE TO THE BLACK HOLE}
\label{section:4.1}
From Model B, we can easily find a steep broken power-law emissivity profile, of which the upper limit of the inner emissivity index is pegged at 10 (taking $q_{\text{in}}$ as an example, the contour plot of $a_{*}$ and $q_{\text{in}}$ as shown in Figure \ref{fig:5}), the outer emissivity index is pegged at 3, and a small break radius is pegged at 6 $R_{\mathrm{g}}$. This phenomenon is common among BHXRBs, which can be explained by a compact corona close to the black hole (e.g., \citealt{Wilkins2012}). This is because, if the corona is far away from the black hole, the light-bending will be weak, and the reflection fraction will be less than 1. Conversely, our results show that the reflection fraction of \texttt{relxillCp} in Model B is $R_{\text {ref}}$=$2.87 \pm 0.84$, which confirms our inference. 

\subsection{EFFECT OF IRON ABUNDANCE}
\label{section:4.2}
In Model B, we fix iron abundance to a standard solar iron abundance, but the reality may be more complex. To test the quantity effect on fitting results caused by variation of iron abundance, we change the iron abundance to its typical value ($A_{\mathrm{Fe}}=5 A_{\mathrm{Fe}, \odot}$) of most sources (\citealt{Javier2018b}), and name it as Model C. For the convenience of comparison, we also list the result of Model C in Table \ref{table:2}. The spin only changes 0.5\% from Model B ($\chi_{\nu}^{2}$=1.063) to Model C ($\chi_{\nu}^{2}$=1.007), while the inclination angle varies 3.7\%. Therefore, for MAXI J1803-298, we conclude that the iron abundance has a negligible effect on our results.

\subsection{EFFECT OF ELECTRON TEMPERATURE OF CORONA}
\label{section:4.3}
Another parameter that may affect the result is the electron temperature of the corona. For Model D, we fix $kT_{\text{e}}$ to its default maximum value (400 keV) and other parameters are the same as Model B. We still list the relevant result in Table \ref{table:2}. We can easily find that the best-fitting parameters did not change significantly, like spin $a_{*}=0.995^{+0.001}_{-0.002}$ ($a_{*}=0.991$$\pm$0.001) and inclination $i= 73.4 \pm 1.3$ degree (70.8 $\pm$ 1.1 degree) in Model D (Model B). Similarly, we can see that the electron temperature of the corona has no significant effect on the spin.

\subsection{EFFECT OF THE CLOUD FAR AWAY FROM THE CENTER BLACK HOLE}
\label{section:4.4}
Though we can not find additional narrow iron line composition from in Figure \ref{fig:3}. To test the contribution of the cold plasma faraway to the spectra, we still add a second non-relativistic reflection Model \texttt{xillverCp} over Model B. The full expression is \texttt{crabcor*constant*TBabs*(diskbb+relxillCp+xillverCp)} (Model E).  Except for the ionization state ($\log \xi $), the reflection fraction ($R_{\text{ref}}$), and the normalization of \texttt{xillverCp} ($N_{\text {xillverCp}}$), we link the parameters of model \texttt{xillverCp} with that of model \texttt{relxillCp}. Since the materials are neutral owing to the farther distance from the central black hole, we fix the $\log \xi =0$ in \texttt{xillvercp}. Moreover, we fix the reflection fraction of \texttt{xillverCp} to -1 and let $N_{\text {xillverCp}}$ be free. To simplify the fitting parameters, we also fix $q_{\text {in }}$, $q_{\text {in}}$ and $R_{\mathrm{br}}$ to the same value in Model B. We list the result in Table \ref{table:2}. Comparing the results between Model E and Model B, it can be seen that the addition of the non-relativistic reflection component does not improve the results, so we prefer not to add the second reflection component that \texttt{xillverCp} for MAXI J1803-298.

\setcounter{figure}{3}
\begin{figure}
    \includegraphics[angle=0, width=\columnwidth]{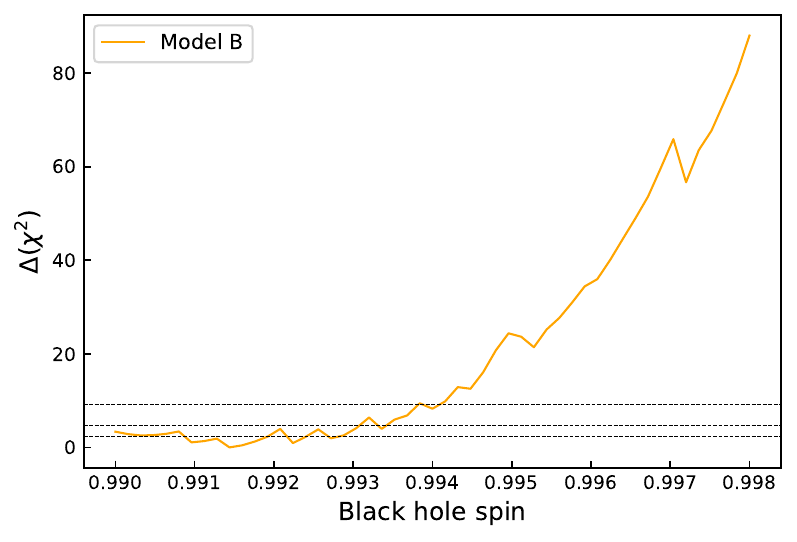}
    \caption{For Model B, the best fit for 50 values of spin from 0.99 to 0.998, by using command \texttt{steppar} in \texttt{XSPEC}.}
   \label{fig:4}
\end{figure}

\setcounter{figure}{4}
\begin{figure}
   \includegraphics[angle=0, width=\columnwidth]{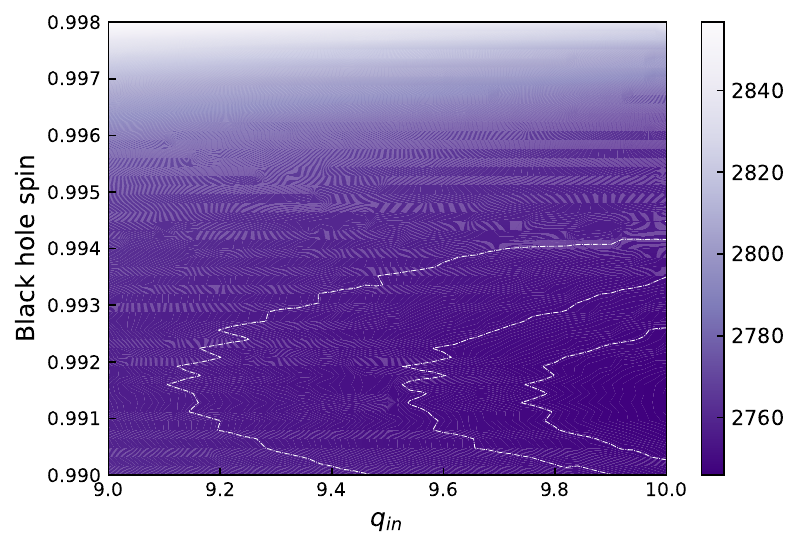}
   \caption{The contour plot of spin and emissivity index of the inner region. The three lines represent 68.3\%, 90\% and 99.7\%, respectively.}
   \label{fig:5}
\end{figure}

\begin{table*}
    \renewcommand\arraystretch{1.5}
    \centering
	 \caption{Properties of \emph{NuSTAR} observation and \emph{NICER} observation}
    \begin{threeparttable}
    \begin{tabular}{ccccccc}
        \hline
        \multirow{2}{*}{\centering ObsID} & \multirow{2}{*}{\centering Instrument/Module} & \multirow{2}{*}{\centering MJD} & \multirow{2}{*}{\centering Start Time} & 
        \multirow{2}{*}{\centering End Time} & \multirow{2}{*}{\makecell{Exposure\\(ks)}} & \multirow{2}{*}{\makecell{Count rates\\ (cts $\rm s^{-1}$)}} \\
        \\
        \hline
        90702318002 & \emph{NuSTAR}/FPMA & 59357 & 2021-05-23 16:24:58.447 & 2021-05-23 20:32:57.207 & 7.2 & 91.73 \\
        90702318002 & \emph{NuSTAR}/FPMB & 59357 & 2021-05-23 16:24:58.471 & 2021-05-23 20:32:57.207 & 7.1 & 83.93 \\
        4202130110 & \emph{NICER}/XTI & 59357 & 2021-05-23 17:24:54.184 & 2021-05-23 20:58:45.184 & 4.9 & 3513 \\
        \hline
    \end{tabular}
    \begin{tablenotes}
        \footnotesize
    \item In columns 1-7, we display the following information: observation ID (ObsID); Instrument/Module; modified Julian date (MJD); start time; end time; exposure time in units of kiloseconds; and net count rates measured at 3.0-79.0 keV (0.7-10.0 keV) for \emph{NuSTAR (NICER)} data in units of cts $\rm s^{-1}$. It should be noted that the information in this table is after the dip time has been removed.
    \end{tablenotes}
    \end{threeparttable}
    \label{table:1}
\end{table*}

\begin{table*}\tiny
    \renewcommand\arraystretch{0.8}
    \centering
	\caption{Best fitting Results for Models}
    \begin{threeparttable}
    \begin{tabular}{cccccccc}
    \hline
     Component  & Parameter\tnote{1} & Description & Model A & Model B &  Model C & Model D & Model E\\
 & & & free $R_{\text{in}}$ & $R_{\text{in}}$=$R_{\text{ISCO}}$ & $A_{\mathrm{Fe}}$=5 & $kT_{\text{e}}=400$ \\
\hline

\tt{constant}              & $C_{\mathrm{FPMA}}$                                                                 &Cross-normalization                                    &	1.095$\pm$0.002                 &1.096 $\pm$0.002                    &	1.096$\pm$0.002              &1.095$\pm$0.002                 &	1.095$\pm$0.002\\
\tt{constant}              & $C_{\mathrm{FPMB}}$                                                                 &Cross-normalization                                    &	1.101$\pm$0.002                 &1.101 $\pm$0.002                    &	1.102$\pm$0.002              &1.101 $\pm$0.002                &	1.101$\pm$0.002\\
\tt{TBabs}                 & $N_{\mathrm{H}}~(\times10^{21}\mathrm{cm}^{-2})$                   &Hydrogen column density                           &     5.00$\pm$ 0.01                      &4.96$\pm$ 0.01                         &4.76$\pm$ 0.01                   &5.06$\pm$ 0.01                    &	4.99$\pm$ 0.01\\
\tt{diskbb}                 & $T_{\mathrm{in}}~(keV)$                                                              &Temperature of disk                                     &     0.78$\pm$0.01                       &0.78$\pm$0.01                         &0.74$\pm$0.01                     &0.79$\pm$0.01                    &	0.78$\pm$0.01\\
\tt{diskbb}                 & $N_{\text {diskbb}}~(\times10^{2})$                                              &Normalization                                              &     5.66$\pm$0.29                        &5.77$\pm$0.30                         &7.67$\pm$0.35                     &5.31$\pm$0.25                    &	5.66$\pm$0.31 \\
\tt{relxillCp}               &     $q_{\text{in}}$                                                                           &Emissivity index in the inner region             & $10_{-0.17}$                               & $10_{-0.17}$                           & $10_{-0.29}$                        &$10^{*}$                              &	$10_{-0.18}$ \\
\tt{relxillCp}               &     $q_{\text{out}}$                                                                         &Emissivity index in the outer region             & $3_{-2.50}$                                 & $3_{-2.50}$                             & $3_{-2.29}$                         &$3^{*}$                                 &	$3_{-0.25}$ \\
\tt{relxillCp}               &     $R_{\text{br}}$~($R_{\text{g}}$)                                                &Break radius                                                & $2.69^{+0.19}_{-0.14}$              &$6_{-2.63}$                              & $6_{-3.13}$                          &$6^{*}$                                 &	$2.63^{+0.16}_{-0.11}$ \\
\tt{relxillCp}               &     $a_{*}$                                                                                      &Black hole spin 		                                &  $0.998^{*}$                                & 0.991$\pm$0.001                    &$0.996^{+0.001}_{-0.002}$  &0.995$\pm$0.001   &	      $0.995^{+0.001}_{-0.002}$ \\
\tt{relxillCp}               &     $i $~($^{\circ}$)                                                                         &Inclination angle                                          & 73.3$\pm$5.5                             &70.8$\pm$1.1	                     &	73.4$\pm$1.6                     &74.7.0$\pm$4.6                    &	73.4$\pm$1.3\\
\tt{relxillCp}               &     $R_{\text{in}} $~($R_{\text{ISCO}}$)                                        &Disk inner radius                                          & 1.08$\pm$0.08                           &$1 ^{*}$                                    &$1 ^{*}$                                 &$1 ^{*}$                                &	$1 ^{*}$\\
\tt{relxillCp}               &     $\Gamma$                                                                                &Photon Index	                                         &	2.33$\pm$0.01                     &	2.33$\pm$0.01                      &	2.28$\pm$0.03                   &2.35$\pm$0.01                     &	2.33$\pm$0.02\\
\tt{relxillCp}               &     $\log \xi $                                                                                   &Ionization state of disk                                &	3.7 $\pm$	0.1                       &	3.7 $\pm$	0.1                        &	4.3 $\pm$	0.3                      &3.6 $\pm$	0.1                  &	3.7 $\pm$	0.1\\
\tt{relxillCp}               &     $A_{\mathrm{Fe}}$~($A_{\mathrm{Fe}, \odot}$)                      &Iron abundance                                            & $1^{*}$                                       & $1^{*}$  	                             & $5^{*}$                                 &$1^{*}$                                  &$1^{*}$ \\
\tt{relxillCp}               &    $ k T_{e}~(\mathrm{keV})$                                                        &Electron temperature of the corona             & $100^{*}$                                    & $100^{*}$  	                             &$100^{*}$                              &$400^{*}$                              &	$100^{*}$\\
\tt{relxillCp}               &     $R_{\text{ref}}$                                                                          &Reflection fraction	                                 &	 3.38$\pm$	1.73             &2.87 $\pm$	0.84                      &	 2.30 $\pm$	1.19           &3.27 $\pm$	0.38                &	3.48 $\pm$	1.02\\
\tt{relxillCp}               &     $N_{\text {relxillCp}}~(\times10^{-2})$                                      &Normalization                                               &	1.39 $\pm$	0.19             &1.39 $\pm$	0.25 	                     &	 1.48 $\pm$	0.28           &1.56 $\pm$	0.12                &	1.36 $\pm$	0.26\\
\tt{xillverCp}             &     $N_{\text {xillverCp}}~(\times10^{-3}) $                                     &Normalization                                               & - 	                                             &  - 	                                     &	-                                          &-                                             &     2.22 $\pm$	1.06\\
\hline
$\chi_{v}^{2}$                          &  ...                                                                                    &  ...                                                               &1.064	                                     &1.063	                                      &1.007               	                  &1.037                                     &1.060 \\
$\chi^{2} ~(\mathrm{d.o.f.})$   &  ...                                                                                    &  ...                                                               &2749.7 (2585) 	                    &2746.7 (2585) 	                     &2601.84 (2585)                     &2683.36 (2588)                      &2739.21 (2584)\\
    \hline
    \end{tabular}
    \begin{tablenotes}
        \footnotesize   
  \item Notes: Except for Model A, Model B/C/D/E are under the assume of $R_{\text{in}}$=$R_{\text{ISCO}}$. Model A/B/C/D is \texttt{crabcor*constant*TBabs*(diskbb+relxillCp)}. Model E is \texttt{crabcor*constant*TBabs*(diskbb+relxillCp+xillverCp)}. The $^{*}$ marker represents the parameter is fixed. The errors are calculated in 68$\%$ confidence interval by \texttt{XSPEC}. 
    \end{tablenotes}
    \end{threeparttable}
    \label{table:2}
\end{table*}

\section{CONCLUSIONS}
\label{section:5}
We perform spectroscopic analysis of the reflection features for the new black hole candidate: MAXI J1803-298, and explore its spin and inclination angle, by analyzing the spectra from \emph{NuSTAR} and \emph{NICER}/{XTI} over the energy bands of 0.7-79.0 keV. Using the X-ray reflection fitting method with different models, we find that:\\
(1) During this observation in SIMS state, MAXI J1803-298 does not have a truncated inner disk.\\
(2) MAXI J1803-298 is a fast-rotating black hole with $a_{*}=0.991$$\pm$ 0.001. \\
(3) MAXI J1803-298 is a high-inclination source with an inclination angle of 70.8$^{\circ}$ $\pm$ 1.1$^{\circ}$, which is consistent with the result (between 60$^{\circ}$-80$^{\circ}$) constrained from its light curve (Figure \ref{fig:1_2}).\\

\section*{Acknowledgements}
We thank the anonymous referee for the constructive comments. Y.F. thanks for the valuable discussions with Prof. Kenji Hamaguchi on extracting \emph{NICER} spectrum. Y.F. thanks Prof. James F. Steiner for the model \texttt{crabcor}. Y.F. also thanks for the useful discussions with Youli Tuo, Prof. Bei You, and Ruican Ma. The software used is provided by the High Energy Astrophysics Science Archive Research Centre (HEASARC), a service of the Astrophysics Science Division at NASA/GSFC and the High Energy Astrophysics Division of the Smithsonian Astrophysical Observatory. This research has made use of the \emph{NuSTAR} Data Analysis Software (NuSTARDAS) jointly developed by the ASI Space Science Data Center (SSDC, Italy) and the California Institute of Technology (Caltech, USA). \emph{NICER} launched on June 3, 2017, which is a 0.2-12 keV X-ray telescope running on the International Space Station (ISS). L. G. is supported by the National Program on Key Research and Development Project (Grant No. 2016YFA0400804), and by the National Natural Science Foundation of China (Grant No. U1838114), and by the Strategic Priority Research Program of the Chinese Academy of Sciences (Grant No. XDB23040100).

\section*{DATA AVAILABILITY}
This article was completed using \emph{NuSTAR} and \emph{NICER} data, which can be obtained from \\
\url{https://heasarc.gsfc.nasa.gov/cgi-bin/W3Browse/w3browse.pl}

\bibliographystyle{mnras}
\bibliography{ref}{}

\bsp	
\end{document}